\documentclass[a4paper,twocolumn,citeautoscript,reprint]{revtex4-1}
\usepackage{amsmath,bm}
\usepackage[dvips]{graphicx,color}
\usepackage[english]{babel}
\usepackage{hyperref}
\usepackage{changepage}
\newcommand{\enquote}[1]{``#1''}

\usepackage[margin=0.8in,bottom=0.75in]{geometry}
\usepackage{caption}
\begin{document}

\title{Stimulated Raman Scattering Imposes Fundamental Limits to the Duration and Bandwidth of Temporal Cavity Solitons}

\author{Yadong Wang}
\email{ywan505@aucklanduni.ac.nz}
\author{Miles Anderson}
\altaffiliation{Current address: \'Ecole Polytechnique F\'ed\'erale de Lausanne (EPFL), CH-1015 Lausanne, Switzerland}
\author{St\'ephane Coen}
\author{Stuart G. Murdoch}
\author{Miro Erkintalo}
\email{m.erkintalo@auckland.ac.nz}

\affiliation{The Dodd-Walls Centre for Photonic and Quantum Technologies, Department of Physics, The University of Auckland, Auckland 1142, New Zealand}

\begin{abstract}
	Temporal cavity solitons (CS) are optical pulses that can persist in passive resonators, and they play a key role in the generation of coherent microresonator frequency combs. In resonators made of amorphous materials, such as fused silica, they can exhibit a spectral red-shift due to stimulated Raman scattering. Here we show that this Raman-induced self-frequency-shift imposes a fundamental limit on the duration and bandwidth of temporal CSs. Specifically, we theoretically predict that stimulated Raman scattering introduces a previously unidentified Hopf bifurcation that leads to destabilization of CSs at large pump-cavity detunings, limiting the range of detunings over which they can exist. We have confirmed our theoretical predictions by performing extensive experiments in several different synchronously-driven fiber ring resonators, obtaining results in excellent agreement with numerical simulations. Our results could have significant implications for the future design of Kerr frequency comb systems based on amorphous microresonators.
\end{abstract}
\maketitle

\indent Temporal cavity solitons (CSs) are pulses of light that can circulate indefinitely in passive driven nonlinear resonators. They were first observed and studied in macroscopic fiber ring cavities, and proposed as ideal candidates for bits in all-optical buffers \cite{Wab1,leo1,Jae1,Jae2}. More recently, studies have demonstrated that temporal CSs can also manifest themselves in monolithic microresonators \cite{herr2,Xue1,Yi1,webb5}, where they play a central role in the generation of stable, low noise, wide bandwidth optical frequency combs \cite{Del1,Kip,Del2}. Because such microresonator frequency combs are very attractive for applications ranging from spectroscopy to telecommunications \cite{gohle1,jones1,hill1,pfeifle1}, there is a growing interest to better understand the dynamics and characteristics of temporal CSs in realistic systems.

\indent \looseness=-1 Temporal CSs are able to persist without changes in their shape or energy thanks to a delicate double-balance \cite{Akhm}. The material Kerr nonlinearity compensates for the solitons' dispersive spreading, while the energy they lose due to intrinsic resonator loss and coupling is replenished through nonlinear interactions with the continuous wave (cw) field driving the cavity. In addition to these fundamental interactions, the precise characteristics of temporal CSs (e.g. duration, temporal profile, center wavelength) can be further influenced by a variety of \enquote{higher-order} effects, such as perturbations induced by higher-order dispersion \cite{Jang,Wang} or avoided mode crossings \cite{Herr1}. In resonators made of amorphous materials, such as silica glass, an effect of particular significance is stimulated Raman scattering (SRS), which causes temporal CSs to be spectrally red-shifted relative to the driving wavelength \cite{Milian,kob,webb5,Yang}. Signatures of such CS self-frequency-shift were first observed experimentally in the context of frequency comb generation in silicon nitride microresonators \cite{kob}, and subsequently in experiments using silica wedge resonators \cite{Yi1} and silica microspheres \cite{webb5}. Indirect time domain signatures have also been reported in macroscopic fiber ring resonators \cite{Ande}.

\indent In addition to shifting the CS center frequency, SRS can also impact on the range of CS existence. Indeed, in a pioneering theoretical work \cite{Milian}, Mili\'an et al. observed that, in the presence of SRS, temporal CSs may not exist over the entire range of parameters where they are expected to do so in the absence of SRS. Yet, the precise fashion in which SRS affects the existence and stability of CSs -- particularly in the context of experimentally realistic systems -- has not to date been investigated. Notably, to the best of our knowledge, no experimental studies have been reported that would demonstrate the impact of SRS on CS existence.

\indent \looseness=-1 In this work, we theoretically and experimentally demonstrate that SRS can significantly restrict the range of CS existence. In particular, we show that, due to SRS, CSs undergo a previously unidentified Hopf bifurcation at large cavity detunings, leading to instability dynamics that limit the range of parameters over which CSs can exist. We show that this new instability sets a fundamental limit for the minimum duration that a CS can possess for given resonator parameters, thereby setting an upper limit for the frequency comb bandwidth that can be achieved in microresonators made of amorphous materials. We have confirmed our theoretical predictions by performing extensive experimental investigations in several different fiber ring resonators; our experimental findings are in very good agreement with numerical simulations.
 
\indent We begin by presenting a general theoretical analysis of the impact of SRS on the existence and stability of temporal CSs. To this end, we analyze the generalized mean-field Lugiato-Lefever equation (LLE) \cite{lugiato9,Coen1,Chembo} that includes the delayed Raman nonlinearity \cite{Milian}. This model is well known to allow for the examination of temporal CS dynamics \cite{Coen1,kob,webb5}. In dimensional form, the equation reads \cite{webb5}:
	\begin{equation}
		\small
		\begin{split}
			t_R\frac{\partial E(t,\tau)}{\partial t}=&\left[ -\alpha-i\delta_\mathrm{0}-\frac{iL\beta_2}{2}\frac{\partial^2}{\partial \tau^2}\right]E+\sqrt{\theta}E_{\text{in}}\\&+i\gamma L\left[(1-f_R)|E|^2+f_Rh_R(\tau)*|E|^2\right]E.
		\end{split}
		\label{LEE1}
	\end{equation}
\noindent Here, $t$ is a \emph{slow} time variable that describes the evolution of the intracavity field envelope $E(t,\tau)$ over consecutive round trips, while $\tau$ is a \emph{fast} time variable defined in a co-moving reference frame that describes the envelope's temporal profile over a single round trip. $t_\mathrm{R}$ is the cavity round trip time, $\alpha$ corresponds to half the total power loss per round trip, $\delta_\mathrm{0}$ is the phase detuning of the driving field $E_\mathrm{in}$ from the closest cavity resonance, $L$ is the cavity round trip length, $\beta_2$ and $\gamma$ are the usual group velocity dispersion and Kerr nonlinearity coefficients, respectively, and $\theta$ is the coupling power transmission coefficient. Finally, $h_\mathrm{R}(\tau)$ is a time-domain response function that characterizes the Raman nonlinearity of the resonator, with $f_\mathrm{R}$ the corresponding Raman fraction. For silica glass, $f_\mathrm{R} \approx 0.18$ and the form of the response function is well known \cite{Stolen,Hollenbeck}. In the calculations that follow, we will assume a silica glass resonator for simplicity, yet emphasize that our general findings are likely to be applicable to arbitrary resonators where CSs exhibit self-frequency-shift.

	\begin{figure}[t]	
		\includegraphics[width=\linewidth]{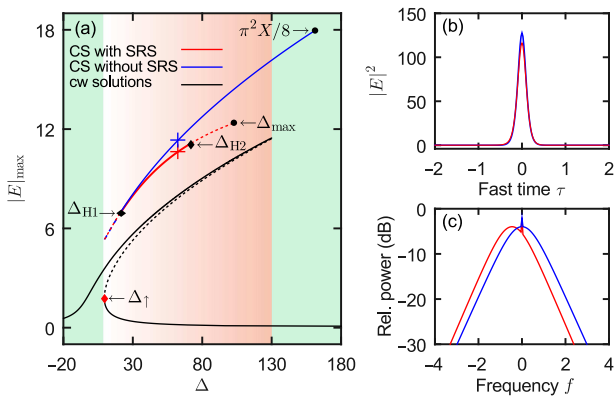}
		\caption{(a) Peak amplitude of the intracavity field, $|E|_\mathrm{max}$, as a function of cavity detuning $\Delta$ for $X=130$. Black curves represent cw solutions, while red and blue curves show CS solutions with and without SRS, respectively. The dashed curves correspond to unstable solutions. (b, c) Temporal (b) and spectral (c) CS profiles for $\Delta = 62$. As in (a), red and blue curves correspond to CS solutions with and without SRS, respectively. The corresponding solutions are marked with crosses in (a). The green and red shaded areas in (a) indicate regions of mono- and bistability of the cw solutions, respectively.}
		\label{fig:Fig1}
	\end{figure}
To gain general insights, we cast Eq.~\eqref{LEE1} into normalized form via the following transformation of variables \cite{leo1,step1}: $\alpha t/t_R\rightarrow t$, $\tau\sqrt{2\alpha/(|\beta_2|L)}\rightarrow\tau$ and $E\sqrt{\gamma L/\alpha}\rightarrow E$. The dimensionless equation reads: 
	\begin{equation}
	\small
	\begin{split}
	\frac{\partial E(t,\tau)}{\partial t}=&\left[ -1-i\Delta+i\frac{\partial^2}{\partial \tau^2}\right]E+S\\&+i\Big[(1-f_\mathrm{R})|E|^2+f_\mathrm{R}\left[\Gamma(\tau,\tau_\mathrm{s})*|E|^2\right]\Big]E,
	\end{split}
	\label{LEE2}
	\end{equation}
\looseness=-1 where we have assumed anomalous dispersion ($\beta_2<0$).  The normalized detuning and driving variables are defined as $\Delta=\delta_\mathrm{0}/\alpha$, $S=E_\mathrm{in}\sqrt{\gamma L\theta/\alpha^3}$, respectively, while the normalized Raman response function $\Gamma(\tau,\tau_\mathrm{s})=\tau_\mathrm{s}h_\mathrm{R}(\tau\tau_\mathrm{s})$ where $\tau_\mathrm{s}=\sqrt{|\beta_2|L/(2\alpha)}$ is the fast time normalization time scale.

\indent In the absence of SRS ($f_\mathrm{R} = 0$), the dynamics and solutions of Eq.~\eqref{LEE2} depend on two parameters only: the cavity detuning $\Delta$ and driving strength $S$ \cite{step1}. However, the delayed nature of the Raman response leads to an additional dependence on the fast time normalization time scale $\tau_\mathrm{s}$. This can be readily understood by noting that the impact of SRS depends on the physical durations (and bandwidths) of the intracavity field features. Indeed, to first order, the convolution term can be approximated as:
	\begin{equation}
	\Gamma(\tau,\tau_\mathrm{s})*|E|^2\approx|E|^2-\frac{T_\mathrm{R}}{f_\mathrm{R}\tau_\mathrm{s}}\frac{\partial|E|^2}{\partial \tau},
	\label{aprox}
	\end{equation}
where $T_\mathrm{R}$ is the Raman time-scale, which is related to the slope of the Raman gain spectrum at zero frequency \cite{miro}. This approximation shows that, to first order, the strength of the Raman term scales as $\tau_\mathrm{s}^{-1}$.

\indent \looseness=-1 To illustrate how SRS affects the stability and existence of CSs, we show in Fig.~\ref{fig:Fig1}(a) the steady-state CS solutions of Eq.~\eqref{LEE2} for a constant driving power $X =|S|^2= 130$ in the presence ($f_\mathrm{R} \approx 0.18$, red curve) and absence ($f_\mathrm{R} = 0$, blue curve) of SRS. Also shown are temporal [Fig.~\ref{fig:Fig1}(b)] and spectral [Fig.~\ref{fig:Fig1}(c)] profiles for a CS at a typical detuning $\Delta = 62$. The solutions were obtained by finding the time-localized steady-state solutions of Eq.~\eqref{LEE2} using a continuation scheme based on the Newton-Raphson method \cite{Coen1}; a normalization time scale $\tau_\mathrm{s} = 1.9$ $\mathrm{ps}$ (corresponding to one of the experiments that will follow) was used for calculations when including SRS. [Note that, in the presence of SRS, the CSs exhibit a time-domain temporal drift, which is accounted for by trivially adjusting the frame of reference of Eq.~\eqref{LEE2}.] To facilitate our discussion, we also show in Fig.~\ref{fig:Fig1}(a) the steady-state cw solutions (black curves) of Eq.~\eqref{LEE2}. The cw solutions exhibit a pronounced bistability, with the middle branch being unconditionally unstable (dashed black curve).

\indent \looseness=-1 As can be seen, for small detunings SRS does not significantly perturb the CS solutions: in both cases (with and without SRS), the solutions exist for detunings above the up-switching point $\Delta_{\uparrow}$, exhibit well known unstable behaviours for small detunings (dashed blue and red curves) \cite{leo2,Ande2,Myu}, and become stable through an inverse Hopf bifurcation ($\Delta_\mathrm{H1}$) as the detuning increases. However, as $\Delta$ increases further, we see a distinct deviation between the solutions obtained in the presence and absence of SRS. This can be understood by noting that the duration of a CS scales inversely with $\sqrt{\Delta}$ \cite{Coen1}: for large detunings the CSs are temporally narrower, and hence spectrally broader, resulting in stronger overlap with the Raman gain spectrum.

\indent \looseness=-1 In the presence of SRS, CSs exhibit lower peak powers, longer durations, and their center frequency is down-shifted [c.f. Fig.~\ref{fig:Fig1}(b) and (c)]. But in addition to perturbing their characteristics, it is evident from Fig.~\ref{fig:Fig1}(a) that SRS also impacts on their range of stability and existence. In particular, we find that, in the presence of SRS, CSs can undergo a second Hopf bifurcation at large detunings [denoted as $\Delta_{\mathrm{H2}}$ in Fig.~\ref{fig:Fig1}(a)], leading to unstable dynamics. Dynamical (split-step) simulations of Eq.~\eqref{LEE2} reveal that these previously unidentified, large-$\Delta$ \emph{unstable} CSs, exhibit behaviors qualitatively similar to those observed for unstable CSs below the first Hopf bifurcation point $\Delta_\mathrm{H1}$ \cite{leo2,Ande2}: for detunings very slightly above $\Delta_\mathrm{H2}$, the CSs exhibit oscillatory behaviour, but as $\Delta$ increases further, they experience an abrupt collapse to the cw state [see Supplementary Material]. The results in Fig.~\ref{fig:Fig1}(a) also show that, in the presence of SRS, the CS solutions cease to exist altogether at a detuning significantly smaller [$\Delta_\mathrm{max}$ in Fig.~\ref{fig:Fig1}(a)] than the theoretical limit of $\pi^2X/8$ observed in the absence of SRS \cite{bara}. We emphasize, however, that $\Delta_\mathrm{max}$ only represents a theoretical upper limit of existence of CS solutions: in practice CSs cannot be sustained for detunings far above $\Delta_{\mathrm{H2}}$ because of the nature of the instability dynamics (see Supplementary Material). At this point we also note that, due to vast differences in the analyzed parameter regions, the instabilities uncovered in Fig.~\ref{fig:Fig1}(a) do not appear straightforwardly related to those described by Mili\'an et al. \cite{Milian}.

	\begin{figure}[t]
		\includegraphics[width=\linewidth]{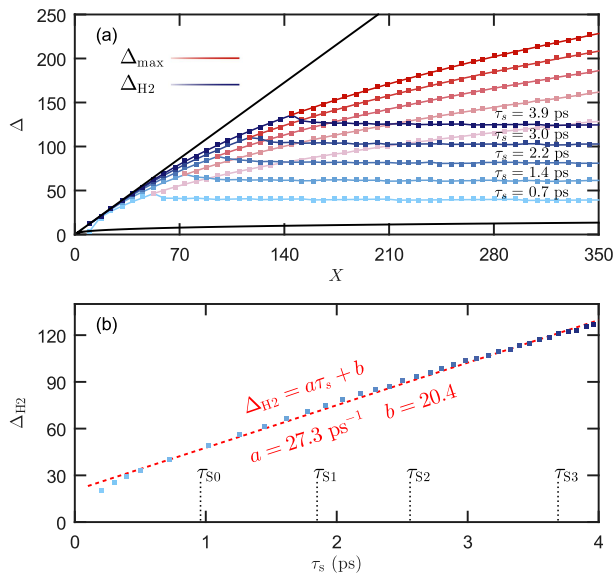}
		\caption{(a) Cavity detunings $\Delta_{\mathrm{H2}}$ and $\Delta_{\mathrm{max}}$, where the CS solution loses its stability and ceases to exist, respectively, as a function  of driving power $X$ and for a variety of $\tau_\mathrm{s}$. The squares correspond to simulated points, while the solid curves are guide to eye. As can be seen, each $\tau_\mathrm{s}$ is associated with a maximum $\Delta_{\mathrm{H2}}$ beyond which CSs cannot remain stable regardless of the driving power. Black curves indicate limits of CS existence in the absence of SRS. (b) Maximum detuning $\Delta_{\mathrm{H2}}$ as a function of the normalization time scale $\tau_\mathrm{s}$. Dashed red curve shows a linear fit. At the bottom of (b), we highlight \emph{four} different normalization time scales corresponding to real resonators. $\tau_\mathrm{S0}$ is from \cite{Yi1} while the others are realized in this work.}
		\label{fig:Fig2}
	\end{figure}
\indent The precise detunings $\Delta_{\mathrm{H2}}$ and $\Delta_{\mathrm{max}}$ (where the CS respectively loses its stability and ceases to exist) depend on the normalization time scale $\tau_\mathrm{s}$ and the driving power $X$. To gain more insight, we have evaluated the CS branches as a function of $\Delta$ [as in Fig.~\ref{fig:Fig1}(a)] for a wide range of $\tau_\mathrm{s}$ and $X$, and extracted $\Delta_{\mathrm{H2}}$ and $\Delta_{\mathrm{max}}$ for each set of parameters. Figure~\ref{fig:Fig2}(a) summarizes our findings. Here we show $\Delta_{\mathrm{H2}}$ (blue curves) and $\Delta_{\mathrm{max}}$ (red curves) as a function of $X$ for five different values of $\tau_\mathrm{s}$. The parameter boundaries, between which CSs can exist in the absence of SRS \cite{leo2}, are also displayed (black curves).

\indent For small driving powers $X$ (and/or large normalization time scales $\tau_\mathrm{s}$), we observe no secondary Hopf bifurcation and the upper limit of CS existence follows closely the expected value of $\pi^2X/8$. However, for larger driving powers (and/or shorter normalization time scales $\tau_\mathrm{s}$), we see clearly that the CS solution loses its stability at large $\Delta$, and that the upper limit of their existence is significantly reduced from the Raman-free values. Surprisingly, while that upper limit $\Delta_\mathrm{max}$ continues to increase with driving power $X$, the upper limit of CS stability (determined by the second Hopf bifurcation point $\Delta_{\mathrm{H2}}$) can be seen to saturate to a constant value that depends exclusively on the normalization time scale $\tau_\mathrm{s}$.

\indent \looseness=-1 The results in Fig.~\ref{fig:Fig2}(a) suggest that, because of SRS, there exist a maximum detuning $\Delta_{\mathrm{H2}}$ above which CSs can no longer remain stable in a given resonator, regardless of the driving power. Furthermore, because of the CS instability dynamics, this detuning also approximates well the upper limit of practical CS existence. As alluded to in Fig.~\ref{fig:Fig2}(a), $\Delta_{\mathrm{H2}}$ depends on the resonator characteristics solely through the normalization time scale $\tau_\mathrm{s}$, and in Fig.~\ref{fig:Fig2}(b) we plot $\Delta_{\mathrm{H2}}$ as a function of $\tau_\mathrm{s}$ as extracted from our calculations. As can be seen, $\Delta_{\mathrm{H2}}$ increases linearly for large $\tau_\mathrm{s}$. For smaller $\tau_\mathrm{s}<0.5\ \text{ps}$, the dependence becomes nonlinear; we speculate this is because the soliton bandwidth becomes comparable with the Raman gain bandwidth.

\indent The observation that SRS imposes a practical upper limit for CS detunings also implies a lower limit for their temporal durations. Indeed, it is well known that, in physical units, the duration of a temporal CS is approximately given by $\tau_0=\tau_\mathrm{s}/\sqrt{\Delta}$ \cite{Coen1,Wab1,herr2}. Thus, in the presence of SRS, the minimum duration that a (stable) CS can possess is $\tau_\mathrm{min}\approx \tau_\mathrm{s}/\sqrt{\Delta_\mathrm{H2}}$. For large $\tau_\mathrm{s} > 1~\mathrm{ps}$ we can further approximate [c. f. Fig.~\ref{fig:Fig2}(b)]:
	\begin{equation}
		\tau_\mathrm{min} = \frac{\tau_\mathrm{s}}{\sqrt{a\tau_\mathrm{s} + b}},
		\label{est}
	\end{equation}
where $a = 27.3\ \mathrm{ps^{-1}}$ and $b = 20.4$ are extracted from the linear fit shown as the dashed red line in Fig.~\ref{fig:Fig2}(b). \emph{This simple linear approximation can be used to estimate the minimum CS duration achievable in silica resonators.} (For other amorphous resonators associated with different Raman responses, the coefficients $a$ and $b$ will likely be different.) To illustrate that this width is consistent with previous experimental findings, we note as an example that the typical duration of CSs observed in a silica wedge resonator is $\tau_0=250\ \mathrm{fs}$ \cite{Yi1}. For the parameters of the resonator, $\tau_\mathrm{S0} \approx 1\ \mathrm{ps}$, yielding a maximum detuning $\Delta_\mathrm{H2} = 47$ and a minimum achievable CS duration $139\ \mathrm{fs}$.

\begin{figure}[t]
	\includegraphics[width=\linewidth]{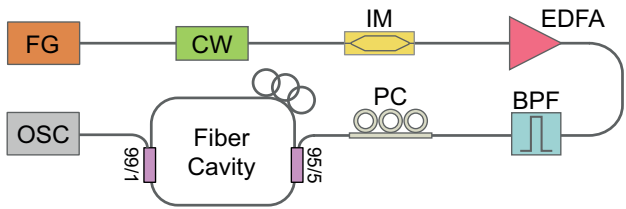}
	\caption{Experimental setup. FG, function generator; CW, cw laser; IM, intensity modulator; EDFA, Erbium-doped fiber amplifier; BPF, bandpass filter; PC, polarization controller; OSC, oscilloscope.}
	\label{fig:Fig3}
\end{figure}

\indent To confirm our theoretical analysis, we have performed experiments using three macroscopic fiber ring resonators associated with different normalization time scales $\tau_\mathrm{s}$. A general schematic of the experimental configurations is depicted in Fig.~\ref{fig:Fig3}. The cavities are made up of single-mode optical fiber (SMF) laid in a ring configuration and closed on themselves by a 95/5 coupler. Each cavity also incorporates a 99/1 tap coupler through which the intracavity dynamics can be monitored in real time. 

\indent Because the cavities contain no other elements, they display very high finesse $\mathcal{F}$. This allows us to reach very high values of normalized driving power $X \propto \mathcal{F}^3$, as required for the study of SRS-induced limits of CS existence [see Fig. 2(a)]. To study the effect of the normalization time scale $\tau_s$, our three different cavities have different round trip lengths of $L = 13\ \mathrm{m}, 25\ \mathrm{m}\ \text{and}\ 50\ \mathrm{m}$, corresponding to normalization time scales $\tau_{\text{S1}}=1.9$ $\mathrm{ps}$, $\tau_{\text{S2}}=2.6$ $\mathrm{ps}$ and $\tau_{\text{S3}}=3.7$ $\mathrm{ps}$, and finesses $\mathcal{F}_\text{1}=77$, $\mathcal{F}_\text{2}=77$ and $\mathcal{F}_\text{3}=69$, respectively.

\indent \looseness=-1 We coherently drive our cavities with flattop nanosecond pulses whose repetition rate is synchronized to the respective cavity round trip time \cite{Coen2,Copie,Ande}. These pump pulses have a duration of $1.2\ \mathrm{ns}$, and they are generated by modulating the output of a narrow linewidth cw laser with a $12$ $\mathrm{GHz}$ intensity modulator. Note that the duration of the pump pulses is sufficiently short to avoid detrimental effects induced by stimulated Brillouin scattering \cite{Agrawal}. Before the pulses are injected into the cavity, they are amplified using an Erbium-doped fiber amplifier (EDFA), and spectrally filtered to remove amplified spontaneous emission. Together with the high cavity finesse, this pulse pumping method allows us to achieve very high normalized driving powers up to $X\approx 200$.

\indent To experimentally explore the limits of CS existence, we linearly tune the cw laser frequency so as to continuously scan the cavity detuning across individual resonances. By simultaneously measuring the cavity output (extracted by the $1\%$ tap coupler) with a fast $12.5\ \mathrm{GHz}$ photodetector, we are able to monitor the intracavity dynamics in real time. This allows us to observe the creation and annihilation of CSs as the detuning is scanned \cite{Luo}, and from the acquired data, we can extract their limits of existence.

\begin{figure}[t]
	\includegraphics[width=\linewidth]{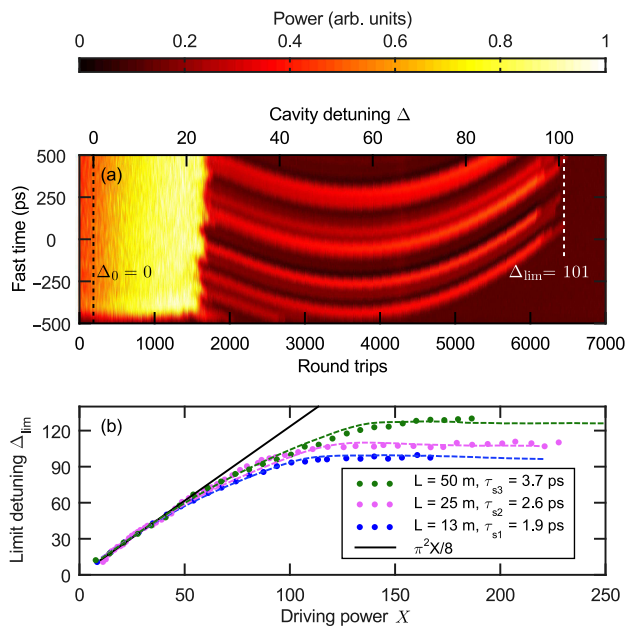}
	\caption{(a) False colour plot showing the intracavity dynamics as the detuning is linearly scanned from 0 to $110$ (top $x$-axis) during about $7000$ round trips (bottom $x$-axis) with a constant driving power $X = 130$. The cavity is 13~m long and has a normalization time constant $\tau_\mathrm{S1} = 1.9\ \mathrm{ps}$. Dashed black and white vertical lines indicate the zero detuning $\Delta_\mathrm{0}$ and the limit detuning $\Delta_\mathrm{lim}$ at which the CSs cease to exist, respectively. (b) Limit detuning $\Delta_\mathrm{lim}$ at which CSs cease to exist as a function of driving power $X$ and for three different normalization time scales $\tau_\mathrm{s}$ as indicated. The circle markers show values extracted from scanning experiments similar to that in (a), while the dashed curves correspond to results from numerical simulations of the LLE. The solid black line shows the theoretical CS existence limit in the absence of SRS, i.e., $\Delta_\mathrm{max} = \pi^2 X/8$.}
	\label{fig:Fig4}
\end{figure}
\indent Figure~\ref{fig:Fig4}(a) shows an example of the measured intracavity dynamics as the cavity detuning $\Delta$ is linearly scanned from 0 to 110 at a constant driving power $X=130$ ($\sim11.3\ \mathrm{W}$ peak power). As is well known, for low detunings the field corresponds to an extended modulation instability pattern which is visible in Fig.~\ref{fig:Fig4}(a) as a solid bright band \cite{Marc}. Out of this chaotic signal, CSs emerge as the detuning increases above $\Delta\approx 30$. The CSs can be seen to exhibit curved trajectories as the detuning increases, which is a known effect of SRS \cite{Ande}. When the detuning reaches $\Delta_\mathrm{lim} \approx 101$, we can see that all the CSs disappear almost simultaneously, indicating the limit detuning beyond which the solitons can no longer be sustained. The observed value is significantly smaller than the theoretical limit $\pi^2X/8 \approx 160$ expected without SRS. Accordingly, the experiment in Fig.~\ref{fig:Fig4}(a) already provides support to our hypothesis: SRS reduces the range of CS existence.

\indent To more comprehensively test our theoretical predictions, we have repeated the above experiment for a wide range of driving powers $X$ and using all three of our resonators to sample different normalization time scales $\tau_\mathrm{s}$. For each experiment, we perform a detuning scan [as in Fig.~\ref{fig:Fig4}(a)], and extract the limit detuning beyond which CSs no longer exist. Our experimental findings are summarized in Fig.~\ref{fig:Fig4}(b). Here the circle markers correspond to experimental data acquired for the different cavities, while the dashed curves correspond to results extracted from realistic numerical simulations of the LLE (the simulations use experimental parameters with no free-running variables). There are several important conclusions to be drawn from Fig.~\ref{fig:Fig4}(b). First, we observe that our numerical simulations are in excellent agreement with experimental findings. Second, in agreement with our theoretical predictions [see Fig.~\ref{fig:Fig2}(a)] the limit detuning $\Delta_\mathrm{lim}$ initially increases with $X$, but eventually saturates to a constant value. This saturation occurs in all of our resonators, but results in different saturated limit detunings due to the different normalization time scales $\tau_\mathrm{s}$. Overall, these measurements confirm of our main hypotheses: SRS limits the range of CS existence, and gives rise to a maximum detuning beyond which CSs cannot exist in a given resonator. 

\indent The experimental results summarized in Fig.~\ref{fig:Fig4}(b) are in excellent qualitative agreement with the theoretical findings presented in Fig.~\ref{fig:Fig2}. However, more careful analysis shows that the agreement is not quantitative, and that our experiments consistently show the upper limit of CSs existence to be greater than the theoretically predicted second Hopf bifurcation point $\Delta_\mathrm{H2}$. For example, the scan of the $13\ \mathrm{m}$ cavity ($\tau_{\mathrm{S1}}=1.9\ \mathrm{ps}$) shown in Fig.~\ref{fig:Fig4}(a) demonstrates a limit detuning $\Delta_\mathrm{lim}=101$, which is higher than the theoretically predicted value $\Delta_\mathrm{H2} = 76$. This discrepancy arises predominantly because the detuning is continuously increased in our experiments. Indeed, the CSs persist briefly even after passing the second Hopf bifurcation point $\Delta_\mathrm{H2}$, and so a continuously increasing detuning naturally leads to overestimation of their existence range. We have confirmed this hypothesis by performing additional experiments where the detuning is initially scanned and then stopped at different points close to the CS existence limit. Results are summarised in supplementary material: they clearly demonstrate that the second Hopf bifurcation point $\Delta_\mathrm{H2}$ represents not only the upper limit of CS stability, but also the practical upper limit of CS existence.

\indent \looseness=-1 To summarize, we have investigated the dynamics of temporal CSs in the presence of stimulated Raman scattering. We have theoretically shown that, due to SRS, temporal CSs can lose their stability through a previously unidentified Hopf bifurcation that occurs for large detunings. Furthermore, we have shown that this instability gives rise to a maximum detuning, which depends solely on the parameters of the resonator, above which CSs cannot exist. Because the duration of temporal CSs scales inversely with $\sqrt{\Delta}$, our theoretical analysis reveals that SRS imposes a fundamental limit on CS durations that can be achieved in resonators made of amorphous materials. We have confirmed our theoretical analyses by performing comprehensive experiments in three different fiber ring resonators. In addition to unveiling a new type of CS instability, our results could significantly impact on the design of frequency comb generators based on microresonators where SRS plays a role, such as silica microspheres \cite{webb5}, wedge resonators \cite{Yi}, fiber-based Fabry-Perot resonators \cite{Obrzud}, or silicon nitride microresonators \cite{kob}.

\begin{acknowledgments}
We acknowledge support from the Marsden Fund of the Royal Society of New Zealand. M. Erkintalo further acknowledges support from the Rutherford Discovery Fellowships of the Royal Society of New Zealand. Y. Wang acknowledges the scholarship from the Dodd-Walls Centre for Photonic and Quantum Technologies. 
\end{acknowledgments}

\pagebreak
\onecolumngrid
\begin{center}
	\textbf{\large Supplementary Information -- Stimulated Raman Scattering Imposes Fundamental Limits to the Duration and Bandwidth of Temporal Cavity Solitons\\}
	
\end{center}

	\begin{adjustwidth}{+1.1cm}{+1.1cm}
		\indent This article contains supplementary information to the manuscript entitled ``\emph{Stimulated Raman Scattering Imposes Fundamental Limits to the Duration and Bandwidth of Temporal Cavity Solitons}''. Specifically, we report on numerical simulations of the mean-field Lugiato-Lefever equation that unveil the dynamical behaviours of unstable cavity solitons in the regime of large pump-cavity detuning. Furthermore, we present additional experimental results that showcase how the second Hopf bifurcation point $\Delta_\mathrm{H2}$ indeed represents a practical upper limit for CS existence.\\
		
	\end{adjustwidth}

\setcounter{equation}{0}
\setcounter{figure}{0}
\setcounter{table}{0}
\setcounter{page}{1}
\makeatletter
\renewcommand{\theequation}{S\arabic{equation}}
\renewcommand{\thefigure}{S\arabic{figure}}
\renewcommand{\bibnumfmt}[1]{[S#1]}
\renewcommand{\citenumfont}[1]{S#1}
\twocolumngrid
\section*{Oscillation Behaviors}
\begin{figure}[th]
	\centering
	\vspace*{+1.3cm}\includegraphics[width = \columnwidth]{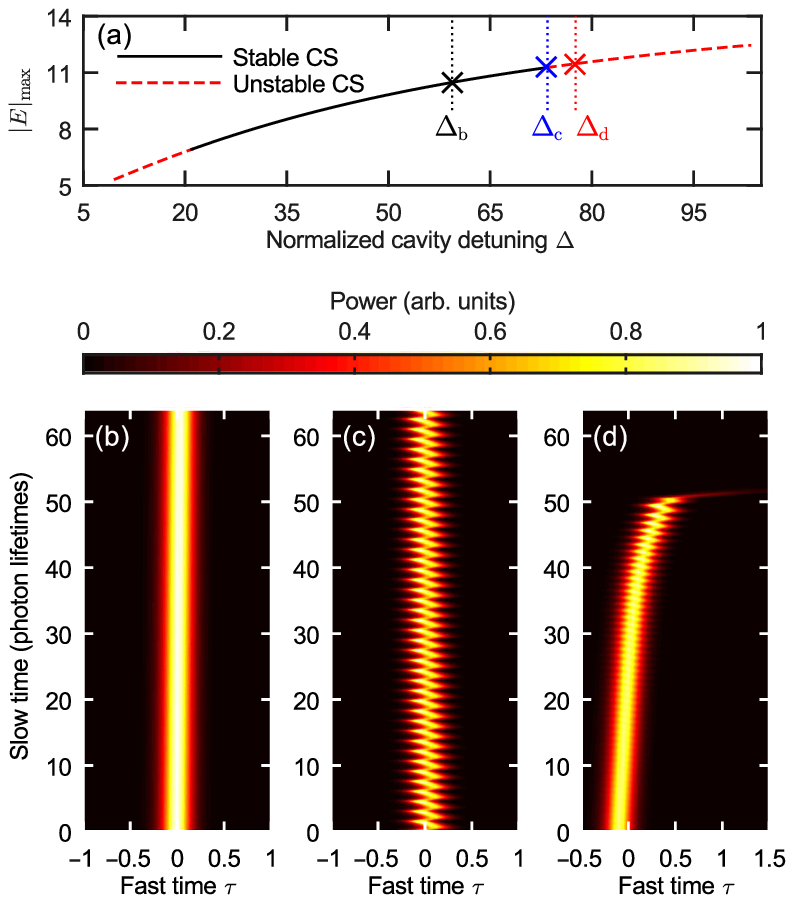}
	\caption{(a) Normalized peak amplitude $|E|_\mathrm{max}$ of the CS solution of the generalized LLE as a function of the normalized cavity detuning $\Delta$ for a constant driving power at $X=130$ and a normalization time scale $\tau_{\mathrm{S1}}=1.9\ \text{ps}$. Stable solutions are indicated with a solid black curve, while dashed red curves indicate unstable solutions. (b-d) Dynamical CS evolutions over $65$ photon lifetimes (about $800$ round trips) at different detunings: (b) stable CS at $\Delta_\mathrm{b}=59.4$, (c) oscillating CS at $\Delta_\mathrm{c}=73.5$ and (d) collapsing CS at $\Delta_\mathrm{d}=79$. Note that the CS evolutions are plotted in a reference frame moving with the pulse.}
	\label{rings}
\end{figure}

\indent We begin by reporting results from split-step simulations of the normalized Lugiato-Lefever equation (LLE, Eq.~\eqref{LEE2} of our main manuscript) that illustrate the CS instability dynamics above the second Hopf bifurcation point $\Delta_\mathrm{H2}$. To this end, we first show, in Fig.~\ref{rings}(a), the normalized CS peak amplitude $|E|_\mathrm{max}$ as a function of the cavity detuning $\Delta$ for a constant driving power $X = 130$ and a normalization time scale $\tau_{\mathrm{S1}}=1.9\ \text{ps}$, obtained using a Newton-Raphson continuation algorithm. (Note that the same data is shown in Fig.~\ref{fig:Fig1}(a) of our main manuscript.) Here, to more clearly distinguish the different regimes, we plot the stable CS solutions as a \emph{solid black} curve, whilst unstable solutions are highlighted with \emph{dashed red} curves. As can be seen, the CSs become unstable at large detunings through a second Hopf bifurcation: the blue cross in Fig.~\ref{rings}(a) indicates the detuning $\Delta_\mathrm{H2}$ at which the Hopf bifurcation (approximately) takes place.

\indent To study how the CS instabilities manifest themselves in the vicinity of the new Hopf bifurcation point $\Delta_\mathrm{H2}$, we have performed split-step simulations of the LLE at several different detunings. In Figs.~\ref{rings}(b)--(d), we show three illustrative examples of the different temporal dynamics observed in our simulations. The detunings $\Delta_\mathrm{b}$,  $\Delta_\mathrm{c}$, and $\Delta_\mathrm{d}$ used in these simulations are highlighted in Fig.~\ref{rings}(a), and their respective values are quoted in the figure caption. We note that the CS evolutions are plotted in the reference frame of the soliton, which deviates from the natural reference frame of the LLE due to the well known temporal drift arising from the interplay of Raman-induced red-shift and group velocity dispersion \cite{Milian2}.

\indent \looseness=-1 Results in Fig.~\ref{rings}(b) are obtained with $\Delta_\mathrm{b} = 59.4$, and they show the typical evolution of a stable CS over $65$ photon lifetimes (corresponding to about 800 round trips in one of our experiments). Despite the presence of SRS, the temporal profile of the CS remains constant. In contrast, at a detuning very close to the second Hopf bifurcation point $\Delta_\mathrm{c}=73.5$, the CS 
starts to oscillate [see Fig.~\ref{rings}(c)], exhibiting a zigzag-type trajectory in the time domain. These oscillations seem to persist indefinitely; however, as the detuning increases further away from $\Delta_\mathrm{H2}$, we find that, following a short period of transient oscillations, the CS instability manifests itself as an abrupt collapse to the cw state [see Fig.~\ref{rings}(d)]. In general, we find that the time it takes for the collapse to develop shortens as the detuning further increases beyond $\Delta_\mathrm{H2}$.This collapse behavior represents the predominant CS instability dynamics above the second Hopf bifurcation point: persistent oscillations [like those shown in Fig.~\ref{rings}(c)] only manifest themselves in the \emph{immediate} vicinity of $\Delta_\mathrm{H2}$, while collapses occur at all other higher detunings. For example, using parameters typical to our experiments (normalization time scale $\tau_{\mathrm{S1}}=1.9\ \text{ps}$ and driving power $X=130$), we find that the CSs start to collapse already when the detuning increases above $\Delta = 74$ (the Hopf bifurcation point $\Delta_\mathrm{H2} \approx 73.5$). This observation clearly indicates that the second Hopf bifurcation point $\Delta_\mathrm{H2}$ can be considered not only the upper of limit of CS stability, but also the practical upper limit of CS existence.

\section*{Scan-and-stop Experiment}
\indent \looseness=-1 As highlighted in our main manuscript, the experimentally measured limit detunings $\Delta_\mathrm{lim}$ at which CSs cease to exist are systematically larger than the predicted values of the second Hopf bifurcation point $\Delta_\mathrm{H2}$. We argued that this discrepancy arises from the fact that the detuning is continuously scanned in our experiments. Here we provide supportive evidence for our explanation by reporting on additional experiments where we stop the detuning scan in the vicinity of the CS collapse point. By stopping the scan at various cavity detunings while assessing whether CSs can continue to persist or not, we are able to refine the experimentally deduced limit detuning.

\begin{figure}[t]
	\centering
	\includegraphics[width = \columnwidth]{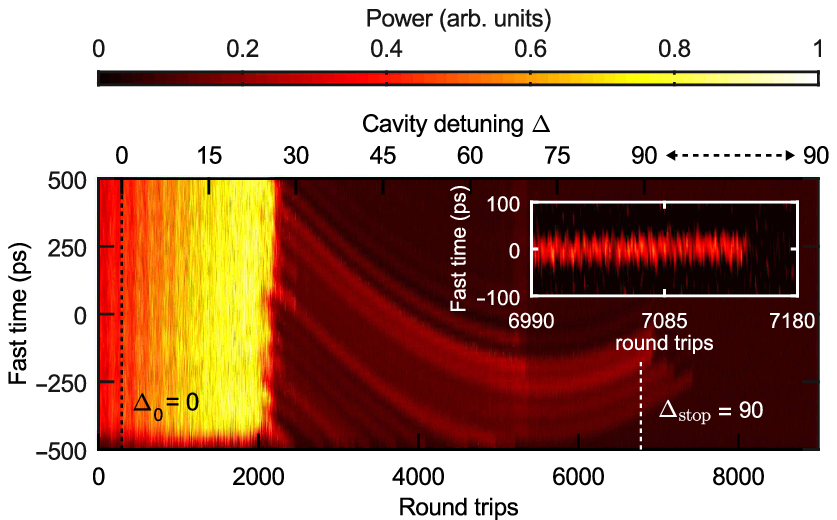}
	\caption{False colour plot of intracavity dynamics during a scan-and-stop experiment performed in our 13 m long fiber cavity with normalization time scale $\tau_{\mathrm{S1}}=1.9$ $\mathrm{ps}$ and normalized driving power $X=130$. The normalized cavity detuning is linearly scanned from  $0$ to $90$ over $6800$ round trips, and then held constant for $2000$ round trips. As can be seen, the CSs collapse after the detuning scan is stopped. The inset is a zoomed-up plot of a single collapsing CS.}
	\label{scanstop}
\end{figure}
\indent \looseness=-1 Out experimental scan-and-stop method is similar to that in ref. \cite{Ande3}. An electrical ramp signal linearly sweeps the wavelength of the cw laser up to a set value, after which point the wavelength is then held constant. By monitoring the intracavity field after the detuning sweep has been stopped, we can readily gauge whether the CSs persist or not; repeating the experiment at several different stop points allows us to refine the limit detuning above which CSs no longer exist. 

\indent Figure~\ref{scanstop} shows a typical example of experimentally measured intracavity dynamics during a scan-and-stop sequence. The cavity used in this experiment is $13$~m long corresponding to a normalization time scale $\tau_{\mathrm{S1}}=1.9$ $\mathrm{ps}$, and the normalized driving power is set to $X = 130$. Using a detuning scan rate similar to that in Fig.~\ref{fig:Fig4}(a) of our main manuscript, we sweep the detuning from $\Delta_0\approx 0$ to $\Delta_\mathrm{stop} = 90$, and then stop the scan and let the cavity evolve freely. As seen in Fig.~\ref{scanstop}, all of the CSs decay within about $1000$ round trips after the detuning scan is stopped. This measurement thus reveals that, for these experimental parameters, the true limit detuning beyond which CSs cannot exist is smaller than $\Delta_\mathrm{stop} = 90$. We note that this value is already smaller (and closer to $\Delta_\mathrm{H2} = 76$) than the value obtained with continuous scanning, namely $\Delta_\mathrm{lim} = 101$. Note that, because of the limited resolution of our photodiode ($\sim60$ ps impulse response), we are unable to fully resolve the CS instability dynamics. Nevertheless, as shown in the inset of Fig.~\ref{scanstop}, some preliminary traces of oscillatory behaviours can be observed before the collapse. 

\begin{figure}[t]
	\centering
	\includegraphics[width = \columnwidth]{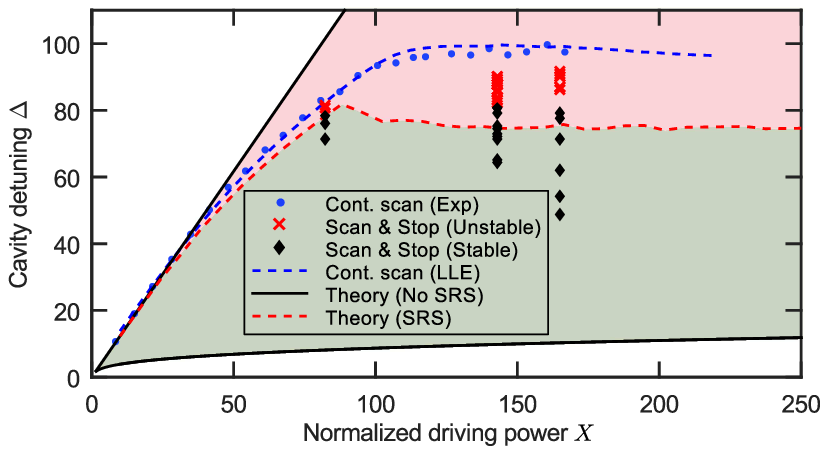}
	\caption{Regions of CS existence and stability for a normalization time scale $\tau_\mathrm{S1}=1.9\ \text{ps}$. Black solid curves show boundaries of CS existence without SRS, blue circle markers show limit detunings $\Delta_\mathrm{lim}$ extracted from measurements with a continuously increasing detuning, dashed blue line shows corresponding results from numerical simulations of the LLE, and dashed red curve shows the theoretically predicted Hopf bifurcation points $\Delta_\mathrm{H2}$. Black diamonds and red crosses depict results from scan-and-stop experiments: diamonds indicate stop-detunings where the CSs persist stably, while the crosses correspond to stop-detunings where the CSs collapse. Grey and pink shaded areas correspond to theoretically predicted regions where the CS solutions are stable or unstable, respectively.}
	\label{Noname}
\end{figure}
\indent To systematically refine the limit value of CS existence, we repeat the scan-and-stop measurement above at several different stop detunings. Furthermore, we have performed these measurements for three different pump powers. The results are summarized in Fig.~\ref{Noname}. Here, the solid black curves show CS existence boundaries in the absence of SRS. The dashed blue curve and circle markers show numerical and experimental data, respectively, of the limit detuning at which the CSs collapse with continuous scanning (same data is shown in Fig.~\ref{fig:Fig4}(a) of our manuscript), and the dashed red curve shows $\Delta_{\mathrm{H2}}$ extracted from Newton-Raphson calculations (same data is shown in Fig.~\ref{fig:Fig2}(a) of our manuscript). New experimental data from our scan-and-stop measurements are shown as black diamonds and red crosses: the former correspond to detuning stop points where the CSs persist stably, while the latter correspond to stop points where the CSs decay (as in Fig.~\ref{scanstop}). This data supports our hypothesis regarding the influence of continuous scanning of the cavity detuning. Indeed, we see that experiments (and simulations) with continuously increasing detuning overestimate the range of CS existence: if the detuning scan is stopped just below the limit detuning found with continuous scanning, the CSs are found to be unstable and eventually collapse. Our experiments further show that the CSs cease to collapse only when the detuning is stopped very slightly above the theoretically predicted second Hopf bifurcation point $\Delta_{\mathrm{H2}}$; below that point the CSs are always found to be stable. 

\indent \looseness=-1 Overall our experimental data strongly confirms our main hypothesis: the second Hopf bifurcation point $\Delta_{\mathrm{H2}}$ -- induced by stimulated Raman scattering -- represents a practical upper limit of CS existence.

\end{document}